\documentclass[letterpaper, 10 pt, conference]{ieeeconf} 

\IEEEoverridecommandlockouts                              

\usepackage{bm}
\usepackage{url}
\usepackage{tikz}
\usepackage{cite}
\usepackage{dsfont}
\usepackage{textcomp}
\usepackage{upgreek}
\usepackage{footnote}
\usepackage{verbatim}
\usepackage{mathrsfs} 
\usepackage{graphicx}
\usepackage{paralist}
\usepackage{multirow}
\usepackage{multicol}
\usepackage{slashbox}
\usepackage{algorithm}
\usepackage{algorithmicx}
\usepackage{algpseudocode}
\usepackage{listings, lstautogobble}
\usepackage[skip=0cm,list=true,labelfont=it]{subcaption}
\algrenewcommand\textproc{} 
\usepackage{amssymb, amsmath, times}
\usepackage[pagebackref=true,breaklinks=true,colorlinks=true, citecolor=green,linkcolor=cyan,urlcolor=blue,bookmarks=true]{hyperref}
\def\BibTeX{{\rm B\kern-.05em{\sc i\kern-.025em b}\kern-.08em
		T\kern-.1667em\lower.7ex\hbox{E}\kern-.125emX}}

\title{\LARGE \bf
The Python LevelSet Toolbox (LevelSetPy)
}

{}

\author{Lekan Molu
\\
\href{https://github.com/robotsorcerer/levelsetpy}{\texttt{https://github.com/robotsorcerer/levelsetpy}}
\thanks{Microsoft Research,
        300 Lafayette Street, NYC
        {\tt\small lekanmolu@microsoft.com}}%
}

\begin{document}

\maketitle
\thispagestyle{empty}
\pagestyle{empty}

\definecolor{light-blue}{rgb}{0.30,0.35,1}
\definecolor{light-green}{rgb}{0.20,0.49,.85}
\definecolor{purple}{rgb}{0.70,0.69,.2}

\newcommand{\lb}[1]{\textcolor{light-blue}{#1}}
\newcommand{\rev}[1]{\textcolor{red}{#1}}

\renewcommand{\figureautorefname}{Fig.}
\renewcommand{\sectionautorefname}{$\S$}
\renewcommand{\equationautorefname}{equation}
\renewcommand{\subsectionautorefname}{$\S$}
\renewcommand{\chapterautorefname}{Chapter}

\newcommand{\cmt}[1]{{\footnotesize\textcolor{red}{#1}}}
\newcommand{\todo}[1]{\textcolor{cyan}{TO-DO: #1}}
\newcommand{\review}[1]{\noindent\textcolor{red}{$\rightarrow$ #1}}
\newcounter{mnote}
\newcommand{\marginote}[1]{\addtocounter{mnote}{1}\marginpar{\themnote. \scriptsize #1}}
\setcounter{mnote}{0}
\newcommand{\ie}{i.e.\ }
\newcommand{\eg}{e.g.\ }
\newcommand{\cf}{cf.\ }
\newcommand{\yes}{\checkmark}
\newcommand{\no}{\ding{55}}

\newcommand{\flabel}[1]{\label{fig:#1}}
\newcommand{\seclabel}[1]{\label{sec:#1}}
\newcommand{\tlabel}[1]{\label{tab:#1}}
\newcommand{\elabel}[1]{\label{eq:#1}}
\newcommand{\alabel}[1]{\label{alg:#1}}

\newcommand{\bull}[1]{$\bullet$ #1}
\newcommand{\argmax}{\text{argmax}}
\newcommand{\argmin}{\text{argmin}}
\newcommand{\mc}[1]{\mathcal{#1}}
\newcommand{\bb}[1]{\mathbb{#1}}

\newcommand{\shmargin}[2]{{\color{magenta}#1}\marginpar{\color{magenta}\raggedright\footnotesize #2}}
\newcommand{\shnote}[1]%
{\textcolor{magenta}{SH: #1}}
\newcommand{\lmnote}[1]%
{\textcolor{orange}{LM: #1}}

\def\tidx{t}
\newcommand{\Note}[1]{}
\renewcommand{\Note}[1]{\hl{[#1]}}  

\def\kau{\mc{K}}
\def\particle{\bm{x}}
\def\materialresponse{\bm{G}}
\def\orthoggroup{{\textit{SO}}(3)}
\def\liegroup{{\textit{SE}}(3)}
\def\liealgebra{\mathfrak{se}(3)}
\def\identity{\bm{I}}
\newcommand{\trace}[1]{\textbf{tr}(#1)}

\def\flock{F}
\def\rot{{R}}
\def\rthree{\bb{R}^3}
\def\reline{\bb{R}}
\def\targetset{\mathcal{L}}
\def\traj{\xi}
\def\ren{\bb{R}^n}
\def\skew{S}
\def\state{\bm{x}}
\def\statex{x}
\def\statey{y}
\def\statez{z}
\def\hot{h.o.t.\ }
\def\lhs{l.h.s.\ }
\def\rhs{r.h.s.\ }
\def\identity{I}
\def\costdiff{\mathbf{\tilde{V}}}
\def\pursuer{\bm{P}}
\def\evader{\bm{E}}
\def\gain{\bm{k}}
\def\control{\bm{u}}
\def\disturb{\bm{w}}
\def\switchcurve{\bm{\gamma}}
\def\valuefunc{\bm{v}}
\def\valu.term{\bm{g}}
\def\lpspace{L^2({\mc{S}}; \mc{F})}
\def\lpdual{L^2({\mc{S}}; \breve{\mc{F}})}
\def\valuetensor{\mathds{V}}
\def\wtensor{\mathds{W}}
\def\valuecore{\mathds{V}^c}
\def\hamfunc{\bm{H}}
\def\hamtensor{\mathds{H}}
\def\uppervalue{\bm{V}^+}
\def\lowervalue{\bm{v}}
\def\upperham{\bm{H}^+}
\def\lowerham{\bm{H}}
\def\hilbertparam{\bm{\phi}}
\def\hilbertcoeff{\bm{\psi}}
\def\hilbertparamspace{\bm{\Phi}}
\def\hilbertcoeffspace{\bm{\Psi}}
\def\hilbertspace{\mathcal{F}}
\def\hilbertdual{\mathcal{S}}
\def\reducedbasis{\Xi_r}
\def\basis{\mathbf{e}}
\def\openset{\Omega}
\def\spatialdomain{\Omega}
\def\timeinterval{I}
\def\liederi{L}
\def\grid{\textbf{g}}
\def\payoff{\bm{\phi}}
\def\Payoff{\bm{\Phi}}

\def\group{\mc{C}}
\def\subgroup{\mc{S}}

\newcommand{\cmark}{\ding{51}}%
\newcommand{\xmark}{\ding{55}}%

\newcommand{\win}{\emoji{window}}
\newcommand{\mac}{\emoji{apple}}
\newcommand{\linux}{\emoji{penguin}}
\newcommand{\opensource}{\emoji{magnifying-glass-tilted-right}}
\newcommand{\commercial}{\emoji{money-bag}}

\definecolor{codegreen}{rgb}{0,0.6,0}
\definecolor{codelblue}{rgb}{0,0.0,0.7}
\definecolor{codegray}{rgb}{0.5,0.5,0.5}
\definecolor{codepurple}{rgb}{0.58,0,0.82}
\definecolor{backcolour}{rgb}{0.95,0.95,0.92}

\newcommand*\greencheck{\textcolor{codegreen}{\ding{52}}}
\newcommand*\redcross{\textcolor{red}{\ding{55}}}

\definecolor{orgred}{rgb}{0.8078,0.4471,0.2314}
\definecolor{darkgreen}{rgb}{0.4157,0.6,0.333}
\definecolor{darkblue}{rgb}{0.0,0.0,0.6}
\definecolor{cyan}{rgb}{0.0,0.6,0.6}
\definecolor{light-gray}{gray}{0.80}
\definecolor{brown}{rgb}{48,19,0}

\definecolor{orange}{rgb}{1,0,0}
\definecolor{grey}{rgb}{0.5,0.5,0.5}
\definecolor{pythoncolor}{rgb}{0.96,0.88,0.76}
\definecolor{joint1color}{rgb}{0.96,0.93,0.89}
\definecolor{others}{rgb}{0.76,0.74,0.82}
\definecolor{salmon}{rgb}{.83,.53,0.53}
\newcommand{\coloropacity}{!65}%
\newcommand{\Hilight}[1]{\makebox[0pt][l]{\color{#1}\rule[-2pt]{0.95\columnwidth}{8pt}}} 

\lstdefinestyle{xmlStyle}{
	basicstyle=\ttfamily\scriptsize,
	columns=fullflexible,
	showstringspaces=false,
	commentstyle=\color{darkblue},
	morecomment=[l]{//},
	numbers=left,                    
	numbersep=10pt, 
	numberstyle=\tiny,
	captionpos=b,
}
\lstset{style=xmlStyle,escapechar=|}

\lstdefinestyle{pythonStyle}{
	basicstyle=\ttfamily\scriptsize,
	columns=fullflexible,
	showstringspaces=false,
	commentstyle=\color{darkblue},
	morecomment=[l]{//},
	numbers=left,                    
	numbersep=10pt, 
	numberstyle=\tiny,
	captionpos=b,	
	autogobble=true
}
\lstset{style=xmlStyle,escapechar=|}

\begin{abstract}
	This paper describes open-source scientific contributions in \textcolor{blue}{python} surrounding the numerical solutions to hyperbolic Hamilton-Jacobi (HJ) partial differential equations viz., their implicit representation on co-dimension one surfaces; dynamics evolution with levelsets; spatial derivatives; total variation diminishing Runge-Kutta integration schemes; and their applications to the theory of reachable sets. They are increasingly finding applications in multiple research domains such as reinforcement learning, robotics, control engineering and automation. We describe the library components, illustrate usage with an example, and provide comparisons with existing implementations. This GPU-accelerated package allows for easy portability to many modern libraries for the numerical analyses of the HJ equations. We also provide a CPU implementation in python that is significantly faster than existing alternatives. 
\end{abstract}

\section{Overview}

The \textit{reliability} 
of the modern automation algorithms that we design 
has become paramount given the dangers that may evolve if nominally envisioned system performance falters. Even so, the need for scalable and faster numerical algorithms in software for \textit{verification} 
and \textit{validation} 
has become timely given the emergence of complexity of contemporary systems. %
The foremost open-source verification software for engineering applications based on Hamilton-Jacobi (HJ) equations~\cite{Kruzkov1970, EvansPDEBook} and levelset methods~\cite{SethianLSBook, LevelSetsBook} is the CPU-based  \texttt{MATLAB}\textregistered-implemented levelsets toolbox~\cite{MitchellLSToolbox}, developed before computing via graphical processing units (GPU) became pervasive. Since then, there has been significant improvements in computer hardware and architecture design, code parallelization algorithms, and compute-acceleration on modern GPUs.  

This  paper describes a python-based GPU-accelerated scientific software package for numerically resolving generalized  discontinuous solutions to Cauchy-type (or time-dependent) HJ hyperbolic partial  differential equations (PDEs). HJ PDEs arise in many contexts including (multi-agent) reinforcement learning, robotics, control theory, differential games, flow, and transport phenomena. We focus on the numerical tools for safety assurance (ascertaining the freedom of a system from harm) in a verification sense in this paper. Accompanying the package are implicit calculus operations on dynamic codimension-one interfaces embedded within  $\mathbb{R}^n$ surfaces, and spatial and temporal discretization schemes for HJ PDEs. Furthermore, we describe explicit integration schemes including the Lax-Friedrichs, Courant-Friedrichs-Lewy (CFL), and total variation diminishing Runge-Kutta (or TVD-RK) conditioning schemes for  HJ Hamiltonians of the form $\hamfunc(\state, \bm{p})$, where $\state$ is the state and $\bm{p}$ is the co-state. Finally, extensions to reachability analyses for continuous and hybrid systems, formulated as optimal control or game theory problems using viscosity solutions to HJ PDEs are described. 

All data transfers to the GPU are based on CuPy~\cite{CUPY} framework. In all, we closely follow the Python Enhancement Proposals (PEP) 8 style guide\footnote{Python PEP 8 style guide: \href{https://peps.python.org/pep-0008/}{\textcolor{codelblue}{peps.python.org/pep-0008/}}}; however, in order not to break readability with respect to the original \texttt{MATLAB}\textregistered code, we err in consistency with the \texttt{MATLAB}\textregistered project layout. The Python package and installation instructions are available on the author's github repository: \href{https://github.com/robotsorcerer/levelsetpy}{\textcolor{codelblue}{\texttt{levelsetpy}}}. The CPU implementation (in Python) is on the  \href{https://github.com/robotsorcerer/levelsetpy/tree/cpu-numpy}{\textcolor{codelblue}{\texttt{ \text{cpu-numpy}}}} tree of the repository.  Extensions to other python GPU programming language are straightforward (as detailed in the CuPy \href{https://docs.cupy.dev/en/stable/user_guide/interoperability.html}{\texttt{\textcolor{codelblue}{interoperability document}}}). While our emphasis is on the resolution of safe sets in a reachability verification context, the applications of this package extend beyond control engineering.

\section{Background and Motivation}
Our interest is in the evolution form of the HJ equation 
\begin{align}
	\lowervalue_t(\state,t) &+ \hamfunc(t; \state, \nabla_x \lowervalue) = 0 \text{   in   } \Omega \times (0, T] 
	\label{eq:ivp}\\
	\lowervalue(\state,t) &= \bm{g}, \text{ on } \partial \Omega \times \{t=T\},  \lowervalue(\state, 0) = \lowervalue_0(\state) \text{ in } \Omega \nonumber
\end{align}
or its convection counterpart 
\begin{align}
	\lowervalue_t &+ \sum_{i=0}^{N} f_i(u)_{\state_i} =0, \,\, \text{ for } t > 0, \state \in \ren, \nonumber \\
	\lowervalue(\state,& 0) = \lowervalue_0(\state), \,\, \state \in \ren
\end{align}
where $\state$ is the state within an open set, $\Omega \subseteq \ren$; $\lowervalue_t$ denotes the partial derivative of the solution $\lowervalue$ with respect to time $t$; the Hamiltonian 
$\bm{H}: (0, T] \times \ren \times \ren \rightarrow \reline$ and $f$ are continuous functions; $\bm{g}, \text{   and   }  \lowervalue_0$ are bounded and uniformly continuous (BUC) functions in $\ren$ --- assumed to be given; and $\nabla_x \lowervalue$ (sometimes represented $\lowervalue_{\state}$) is the spatial gradient of $\lowervalue$. 

Solving problems described by \eqref{eq:ivp} under appropriate boundary and/or initial conditions using the method of characteristics is limiting as a result of crossing characteristics~\cite{Crandall1983viscosity}. In the same vein, global analysis is virtually impossible owing to the lack of existence and uniqueness of  solutions $\lowervalue \in C^1(\Omega) \times (0, T]$ even if $\hamfunc$ and $\bm{g}$ are smooth~\cite{Crandall1983viscosity}.  The method of ``vanishing viscosity", based on the idea of traversing the limit as $\delta \rightarrow 0$ in the hyperbolic equation \eqref{eq:ivp} 
allows generalized (discontinuous) solutions~\cite{Evans1984} whereupon if $\lowervalue \in W_{loc}^{1,\infty}(\Omega) \times (0, T]$ and $\hamfunc \in W_{loc}^{1,\infty}(\Omega)$, one can lay claim to strong notions of  general existence, stability, and uniqueness to BUC solutions $\lowervalue^\delta$ of the  (approximate) viscous Cauchy-type HJ equation 
\begin{align}
	\lowervalue_t^\delta &+ \hamfunc(t; \state, \nabla_x \lowervalue^\delta) - \delta \Delta \lowervalue^\delta = 0 \text{   in   } \Omega \times (0, T] 
	\label{eq:ivp-viscous} \\
	\lowervalue^\delta(\state,&t) = \bm{g}, \text{ on } \partial \Omega \times \{t=T\}, \lowervalue^\delta(\state, 0) = \lowervalue_0(\state) \text{ in } \Omega  \nonumber
\end{align}
in the class $\text{BUC}(\Omega \times [0, T]) \cap C^{2,1}(\Omega \times (0, T])$ \ie continuous second-order spatial and first order time derivatives for all time $T<\infty$. 
Crandall and Lions~\cite{CrandallTwoApprox} showed that $|\lowervalue^\delta(\state,t) - \lowervalue(\state,t)| \le k\sqrt{\delta}$ for a small $k>0$. 
Throughout the rest of this paper, we are concerned with \textit{generalized} viscosity solutions of the manner described by \eqref{eq:ivp-viscous}.

Reachability concerns evaluating the \textit{decidability} of a dynamical system's evolution of trajectories throughout a state space. Decidable reachable systems are those where one can compute all states that can be reached from an initial condition in \textit{a finite number of steps}. For $\inf$-$\sup$ or $\sup$-$\inf$ optimal control problems~\cite{LygerosReachability}, the Hamiltonian is related to the \textit{backward} reachable set~\cite{Mitchell2020} of a dynamical system.  Mitchell~\cite{Mitchell2005} connected techniques used in levelset methods to reachability analysis in optimal control, essentially showing that the zero-levelset of the differential zero-sum two-person game in an  HJ-Isaacs (HJI) setting~\cite{Isaacs1965,Evans1984} constitutes the safe set of a reachability problem~\cite{LygerosReachability}.  We refer interested readers on the technicalities of the theory to Mitchell's paper~\cite{Mitchell2005}. 

The well-known \texttt{LevelSet Toolbox}~\cite{MitchellLSToolbox} is the consolidated \texttt{MATLAB}\textregistered\ package that contains the grid methods, boundary conditions, time and spatial derivatives, integrators and helper functions. While Mitchell motivated the execution of the toolkit in \texttt{MATLAB}\textregistered\ based on its expressiveness, modern high-dimensional research and engineering problems 
often render the original package limiting in computational scalability given its single computer processor implementation, lack of interoperability with many modern computing and scripting languages such as \texttt{Numpy}, \texttt{Scipy}, \texttt{PyTorch} and their variants. In this regard, we revisit the major algorithms necessary for implicit surface representation for HJ PDEs, re-write the spatial, temporal, and monotone difference schemes in Python, and accelerate these schemes on modern GPUs via CuPy~\cite{CUPY}. 
%
%
\noindent Our \textbf{contributions} are as follows:
\begin{enumerate}
	\item we describe the levelset python toolkit, starting with the common implicit surfaces that are used as initial conditions to represent $\lowervalue(\state, t)$; 
	\item we describe our implementations of the upwinding spatial derivative, temporal discretization via method of lines schemes based on (approximate) total variation diminishing (TVD) Runge-Kutta (RK), and stabilizing Lax-Friedrichs schemes for multidimensional monotone Hamiltonians of HJ equations or scalar conservation laws; 
	\item we then conclude with a representative example, namely, the barrier surface for two adversarial rockets traveling on an $xz$-plane. Further examples with execution clock times abound on the online code repository and in this article's \href{https://scriptedonachip.com/downloads/Papers/LevPy.pdf}{journal submission version}.
\end{enumerate}
%

The rest of this paper is structured as follows. 
We describe the geometry of (and Boolean operations on) implicit function representations of continuous-time value functions described by \eqref{eq:ivp} using Cartesian grids in section \ref{sec:implicits}. Spatial derivatives to scalar conservation laws are discussed in section \ref{sec:upwinding}, and temporal discretization schemes for these conservation laws follow thereafter. In section \ref{sec:results} we formulate a didactic two-rockets game and show how to define the  numerical safe backward reachable sets and tubes amenable to HJ PDEs in a  geometrical verification framework. 
We conclude the paper in section \ref{sec:conclude}. Additional \texttt{python} examples, \texttt{jupyter} notebooks, and representative problems are provided in the online package.

\section{The LevelSetPy Python Package}
\label{sec:implicits}
Let us now describe how solutions to the HJ in \eqref{eq:ivp} and \eqref{eq:ivp-viscous} are constructed in our software package. 
\subsection{Geometry of Implicit Surfaces and Layouts}
\begin{figure*}[tb!]
	\centering
	\begin{minipage}[b]{.23\textwidth}
		\includegraphics[width=\textwidth]{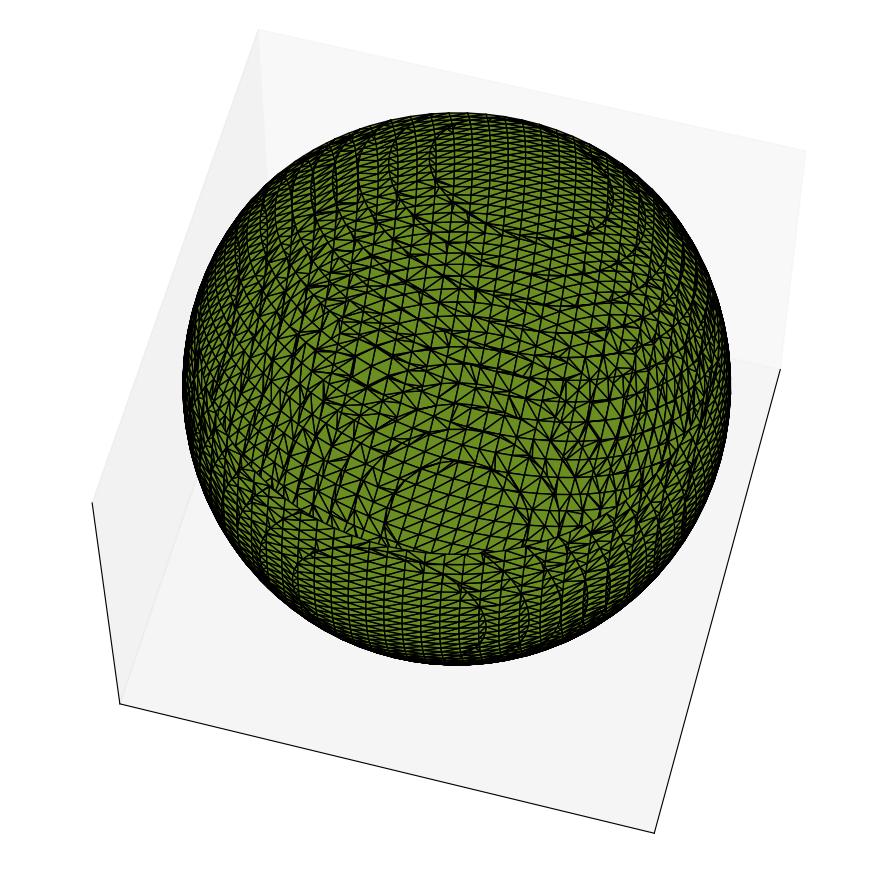}
	\end{minipage}
	\begin{minipage}[b]{.23\textwidth}
		\includegraphics[width=\textwidth]{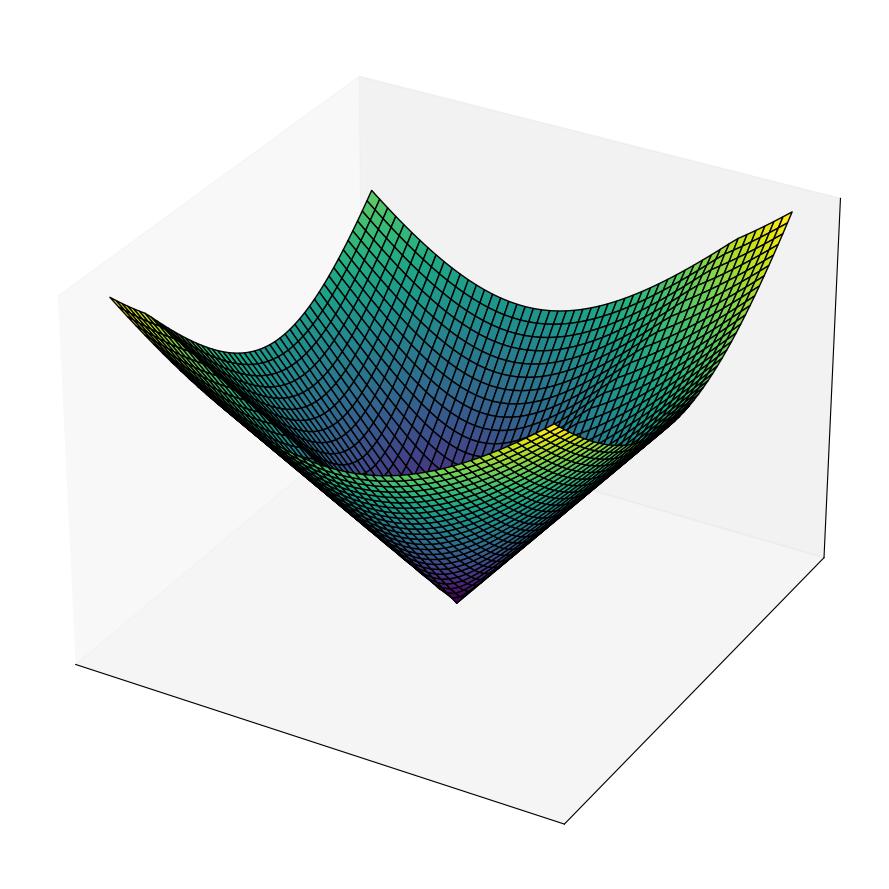} 
	\end{minipage}	
	%
	\begin{minipage}[b]{.23\textwidth}
		\includegraphics[width=\textwidth]{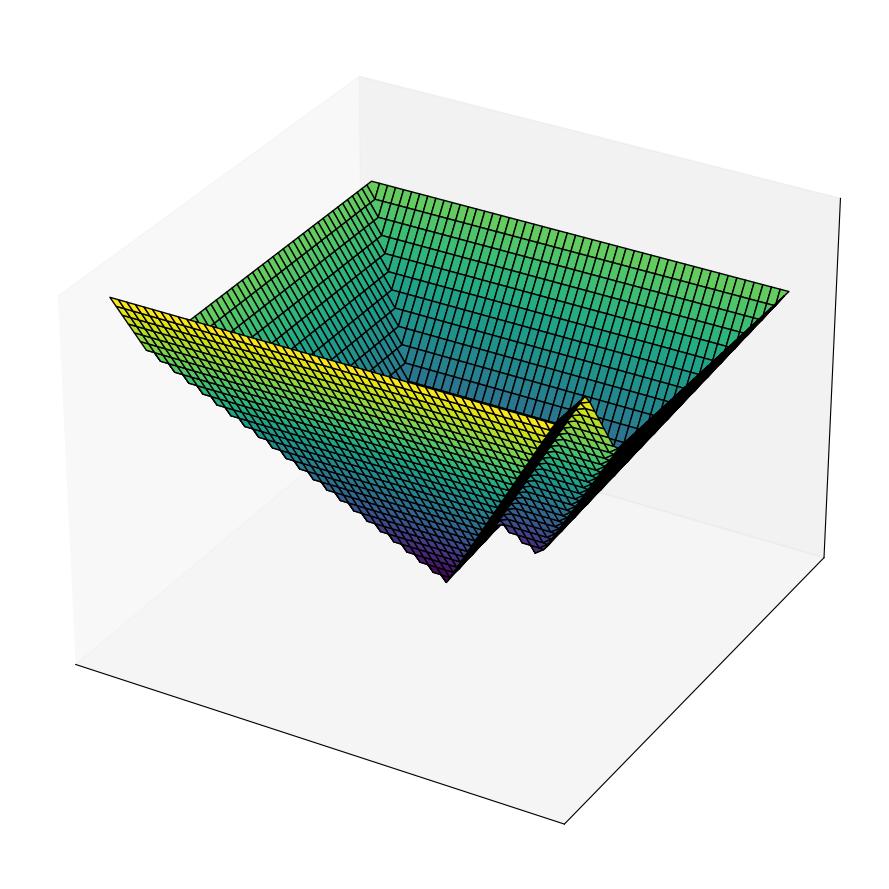} %
	\end{minipage}
	%
	\hfill 
	\begin{minipage}[b]{.23\textwidth}
		\includegraphics[width=\textwidth]{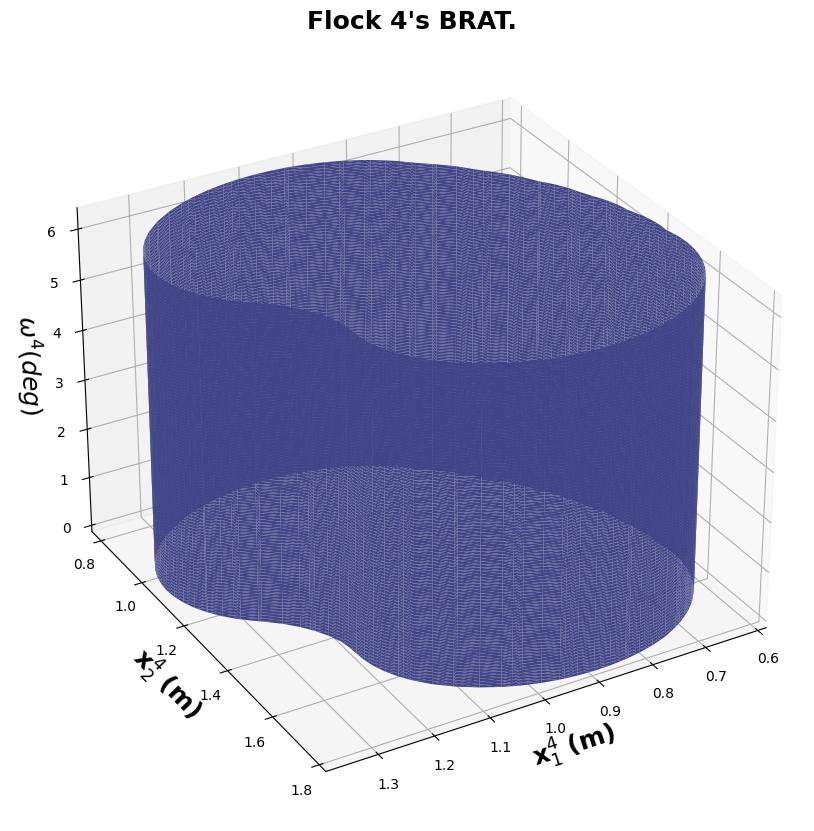}
	\end{minipage}
	%
	%
	\caption{\footnotesize{Implicitly constructed geometric shapes in our library: 
		\textbf{(a)}  a sphere on a 3D grid; \textbf{(b)} union of two 3D spheres implicitly constructed on a  2D grid; \textbf{(c)} the union of rectangles on a 2D grid; \textbf{(d)} the union of multiple cylinders on a 3D grid. 
	}}
	\label{fig:implicit_3D}	
\end{figure*}

As stated before, solutions to the HJ equation \eqref{eq:ivp} are implicitly represented on co-dimension one surfaces in $\ren$. We discuss implicit surface functions' contruction in levelsetpy, CPU memory, and GPU transfers. Throughout, links to \texttt{API's}, \texttt{routines}, and \texttt{subroutines} are hyperlinked and highlighted in \textcolor{codelblue}{\texttt{blue text}} and we use code snippets in Python to illustrate API calls when it's convenient. 

Implicit interfaces are typically isocontours of some function, $f(x)$ --- attractive as they require fewer points to construct a function than explicit representations. The zero isocontour (or levelset)  of a reachable optimal control problem is equivalent to the safety set or backward reachable tube~\cite{Mitchell2001}; and for a differential game, it is the \textit{usable part's boundary} of the \textit{barrier surface} between the \textit{capture} and \textit{escape zones} for all trajectories that emanate from a system.


\subsection{Grids Layout}
Fundamental to implicit surface representations are Cartesian grids in our library. Packages that implement `grid' data structures are in the folder \href{https://github.com/robotsorcerer/levelsetpy/blob/cupy/levelsetpy/grids}{\textcolor{codelblue}{\texttt{grids}}}. A grid $\grid$ is \href{https://github.com/robotsorcerer/levelsetpy/blob/cupy/levelsetpy/grids/create_grid.py}{\textcolor{codelblue}{\texttt{created}}} by specifying minimum and maximum axes bounds, $[\grid_{min}, \grid_{max}]$, along every Cartesian coordinate axes $n$ (see lines 3 and 4) of \autoref{lst:createGrid}; a desired number of discrete points $N$ is passed to the grid data structure -- specifying the number of grid nodes and the grid spacing in each dimension as (line 5) listed in \autoref{lst:createGrid}.  On line 7, the grid data structure is constructed and all input  parameters to the API are checked for consistency. 

\begin{minipage}{0.95\columnwidth} 
	\centering
	\begin{lstlisting}[caption={Creating a three-dimensional grid.},label={lst:createGrid},style=pythonStyle, language=python]
		|\Hilight{pythoncolor\coloropacity}| from math import pi
		|\Hilight{pythoncolor\coloropacity}| import numpy as np
		|\Hilight{pythoncolor\coloropacity}| gmin = np.array((-5, -5, -pi)) // lower corner 
		|\Hilight{pythoncolor\coloropacity}|  gmax = np.array((5, 5, pi))   // upper corner 
		|\Hilight{pythoncolor\coloropacity}|   N = 41*ones(3, 1) // number of grid nodes
		|\Hilight{pythoncolor\coloropacity}|  pdim = 3; // periodic boundary condition, dim 3
		|\Hilight{pythoncolor\coloropacity}|   g = createGrid(gmin, gmax, N, pdim)
	\end{lstlisting}
\end{minipage}
A grid \href{https://github.com/robotsorcerer/levelsetpy/blob/cupy/levelsetpy/grids/process_grid.py}{\textcolor{codelblue}{\texttt{data structure}}}, $\grid$, (implemented in \autoref{lst:createGrid}) has the following fields:
\begin{inparaenum}[(i)]
	\item discretized nodes of the state(s) $\state$ in  \eqref{eq:ivp-viscous}, denoted as 1-D vectors $\grid.vs$; 
	\item given the 1-D vectors $\grid.vs$, an $n$-dimensional array of coordinates over $n$-dimensional grids is computed with matrix-based indexing; this generates a mesh for all state nodal points on the grid $\grid.xs$ as a list across all the dimensions of the grid;
	\item grid dimension $\grid.dim$, denoting the  number of Cartesian axes needed for representing the state $\state$\footnote{This parameter is useful when computing signed distance to every nodal point on the state space in the implicit representation of $\lowervalue$};
	\item boundary conditions of the relavant HJ equation to be solved are grafted in by populating the corresponding grid dimension with ghost cells (to be introduced shortly).
\end{inparaenum}

\subsection{Implicit Surface Representations: Levelsets}
We treat coordinates as functional arguments using a fixed levelset of continuous function $\lowervalue: \reline^n \rightarrow \reline$. We use signed distance functions to represent the dynamics throughout. When the signed distance function is not numerically possible, we describe where the implicit surface representations are smeared out in every routines' \texttt{documentation}. The query points for moving interfaces are grid point sets of the computational domain  described by implicit geometric primitives such as spheres, cylinders, ellipsoids and even polyhedrons such as icosahedrons. All of these are  contained in the folder \href{https://github.com/robotsorcerer/levelsetpy/blob/cupy/levelsetpy/initialconditions/ellipsoid.py}{\textcolor{codelblue}{\texttt{initialconditions}}} on our project page. 

The zero levelset of an implicit surface $\lowervalue(\state,t)$ is defined as $\Gamma = \{\state: \lowervalue(\state, t)=0\}$ on a grid $G\in \ren$, where $n$ denotes the number of dimensions, and the representation of $\Gamma$ on $G$ generalizes a row-major layout.  An example representation of an ellipsoid on a three-dimensional grid is illustrated in \autoref{lst:zerolevelsets}.

\begin{minipage}{0.95\columnwidth} 
	\begin{lstlisting}[caption={An \href{https://github.com/robotsorcerer/levelsetpy/blob/cupy/levelsetpy/initialconditions/ellipsoid.py}{\texttt{ellipsoid}} as a signed distance function.},label={lst:zerolevelsets},style=pythonStyle, language=python]
		|\Hilight{salmon\coloropacity}| e = (g.xs[0])**2 // ellipsoid nodal points
		|\Hilight{salmon\coloropacity}|  e += 4.0*(g.xs[1])**2
		|\Hilight{salmon\coloropacity}|   if g.dim==3:
		|\Hilight{salmon\coloropacity}\hspace{0.2cm}| \texttt{e} += (9.0*(grid.xs[2])**2) 
		|\Hilight{salmon\coloropacity}|  e -=  radius  // radius=major axis of ellipsoid
	\end{lstlisting}
\end{minipage}

\subsection{Calculus on Implicit Function Representations}

Geometrical operations on implicitly defined functions carry through in the package as follows. Let $\lowervalue_1(\state)$ and $\lowervalue_2(\state)$ be two signed distance representations, then the union of the interior of both is simply $\min(\lowervalue_1(\state), \lowervalue_2(\state))$ (illustrated in \autoref{fig:implicit_3D}b and d). The intersection of two signed distance functions' interiors is generated by $\max(\lowervalue_1(\state), \lowervalue_2(\state))$ (illustrated in \autoref{fig:implicit_3D}c). The complement of a function is found by negating its signed distance function \ie $-\lowervalue(\state)$. The resultant function as a result of the subtraction of the interior of one signed distance function $\lowervalue_2$ from the another one, say, $\lowervalue_1$ is defined $\max(\lowervalue_1(\state), -\lowervalue_2(\state))$. All of these are implemented in the module \href{https://github.com/robotsorcerer/levelsetpy/blob/cupy/levelsetpy/initialconditions/shape_ops.py}{\textcolor{codelblue}{\texttt{shapeOps}}}.
%
%
%
%
%

\section{Spatial Discretization: Upwinding}
\label{sec:upwinding}
In this section, we discuss higher-order upwinding schemes that mimic high-order essentially non-oscillatory (ENO)~\cite{OsherShuENO} schemes for computing the spatial derivatives $\lowervalue_x$ for the numerical viscosity solutions to levelset PDEs of the \textit{Eulerian form} introduced in \eqref{eq:levelset}. Routines for procedures herewith described are in the folder \href{https://github.com/robotsorcerer/levelsetpy/tree/cupy/levelsetpy/spatialderivative}{\texttt{spatialderivative}}. Using the Eulerian form of the levelset equation,
\begin{align}
	\lowervalue_t + \bm{F} \cdot \nabla \lowervalue = 0
	\label{eq:levelset}
\end{align} 
where $\bm{F}$ is the speed function, the implicit function representation $v_t$ (see \S \ref{sec:implicits}) is used both to denote and evolve the interface. Suppose that the interface speed $\bm{F}$ is a three-vector $[f_x, f_y, f_z]$ on a three-dimensional Cartesian grid, expanding \eqref{eq:levelset} the evolution of the implicit function on the zero levelset yields the Eulerian form 
\begin{align}
	\lowervalue_t + f_x \lowervalue_x + f_y \lowervalue_y + f_z \lowervalue_z = 0
	\label{eq:upwind_advance}
\end{align}
of the interface evolution given that the interface encapsulates the implicit representation $\lowervalue$. In our implementation, we define $\lowervalue$ throughout the computational domain $\Omega$. 

Let us first  construct the general form of a spatial upwinding scheme \ie
\begin{align}
	D^-\lowervalue = \dfrac{\partial \lowervalue}{\partial \state} \approx \frac{\lowervalue_{i+1} - \lowervalue_i}{\Delta \state}, D^+\lowervalue \approx \frac{\lowervalue_{i} - \lowervalue_{i-1}}{\Delta \state},
	\label{eq:fwd_bck_differencing}
\end{align}
where $\lowervalue$  and its speed $\bm{F}$ are defined over a domain $\openset$ (this is the Cartesian grid in our representation). Using the forward Euler method, the levelset equation \eqref{eq:upwind_advance} becomes
%
$	(1/\Delta t) \cdot {\lowervalue^{n+1}-\lowervalue^n}+ f_x^n \lowervalue_x^n + f_y^n \lowervalue_y^n + f_z^n \lowervalue_z^n = 0.$

Now, suppose that we are on a one-dimensional surface and around a grid point $i$. Given that $f^n$ may be spatially varying, the foregoing equation  becomes
\begin{align}
	\dfrac{\lowervalue^{n+1}_i-\lowervalue^n_i}{\Delta t}+ f^n_i(\lowervalue_x)^n_i  = 0
	\label{eq:upwinding_basic}
\end{align}
where $(\lowervalue_x)_i$ denotes the spatial derivative of $\lowervalue$ w.r.t $x$ at the point $i$. 
We now discuss the specific implementations of the upwinding schemes.

\subsection{First-order accurate upwinding discretization} 
If $f_i > 0$ in \eqref{eq:upwinding_basic}, the values of $\lowervalue$ are traversing from left to right so that in order to update  $\lowervalue$ at the end of the next time step, we must look to the left (going by the method of characteristics~\cite[\S 3.1]{LevelSetsBook}) and vice versa if $f_i < 0$. We therefore follow the standard upwinding method by using \eqref{eq:fwd_bck_differencing}: we approximate $\lowervalue_x$ with $D^-\lowervalue$ whenever $f_i > 0$ and we approximate $\lowervalue_x$ with $D^+\lowervalue$ whenever $f_i < 0$. No approximation is needed when $f_i=0$ since $f_i(\lowervalue_x)_i$ vanishes. This scheme is accurate within $O(\Delta \state)$ given the first order accurate approximations $D^- \lowervalue$ and $D^+\lowervalue$. We have followed the naming convention in \cite{MitchellLSToolbox} and in our \href{https://github.com/robotsorcerer/levelsetpy/tree/cupy/levelsetpy/spatialderivative}{\textcolor{codelblue}{\texttt{spatialderivative}}} folder, we name this function \href{https://github.com/robotsorcerer/levelsetpy/tree/cupy/levelsetpy/spatialderivative/upwind_first_first.py}{\textcolor{codelblue}{\texttt{upwindFirstFirst}}}.

\subsection{ENO Polynomial Interpolation of Solutions}
Using a divided differencing table, essentially non-oscillatory (ENO) polynomial interpolation of the discretization~\cite{OsherShuENO} of the levelset equation is known to generate improved numerical approximations to $D^-\lowervalue$ and $D^+\lowervalue$. Suppose that we choose a uniform mesh discretization $\Delta \state$. Define the zeroth divided differences of $\lowervalue$ at the grid nodes $i$ as
%
$D_i^0 \lowervalue = \lowervalue_i,$
%
and the first, second, and third order divided differences of $\lowervalue$ as the midway between grid nodes \ie
\begin{subequations}
	\begin{align}
		D_{i+1/2}^1 \lowervalue &= \dfrac{D_{i+1}^0\lowervalue - D_i^0 \lowervalue}{\Delta \state}, \,\, D_{i}^2 \lowervalue = \dfrac{D_{i+1/2}^1\lowervalue - D_{i-1/2}^1 \lowervalue}{2\Delta \state}, \\
		D_{i+1/2}^3 \lowervalue &= \dfrac{D_{i+1}^2\lowervalue - D_i^2 \lowervalue}{3\Delta \state}.
	\end{align}
\end{subequations}

Then, an essentially non-oscillating polynomial of the form 
\begin{equation}
	\lowervalue(\state) = Q_0(\state) + Q_1(\state) + Q_2(\state) + Q_3(\state)
	\label{eq:upwinding_approx}
\end{equation}
can be constructed. In this light, the backward and forward spatial derivatives of 
$\lowervalue$ w.r.t $\state$ at grid node $i$ is found in terms of the derivatives of the coefficients $Q_i(\state)$ in the foregoing \ie 
\begin{align}
	\lowervalue_x(\state_i) = Q_1^\prime(\state_i) +  Q_2^\prime(\state_i) + Q_3^\prime(\state_i).
\end{align} 

Define $k=i-1$ and $k=i$ for $\lowervalue_x^-$ and $\lowervalue_x^+$ respectively. Then the first order (\ie first-order upwinding) accurate polynomial interpolation is essentially 
\begin{align}
	Q_1(\state) = (D^1_{k+1/2} \lowervalue)(\state-\state_i),\,\, Q_1^\prime(\state_i) = D^1_{k+1/2} \lowervalue.
\end{align}

We follow Osher and Fedkiw's recommendation~\cite{LevelSetsBook} in avoiding interpolating near large oscillations in gradients. Therefore, we choose a constant $c$ such that 
\begin{align}
	c = \begin{cases}
		D^2_k\lowervalue \qquad \text{   if   } | D^2_k \lowervalue| \le | D^2_{k+1} \lowervalue | \\
		D^2_{k+1} \lowervalue \qquad \text{  otherwise  } 
	\end{cases}
\end{align} 
so that 
%
\begin{align}
	Q_2(\state) =c(\state-\state_k)(\state-\state_{k+1}),\,	Q_2^\prime(\state_i) = c(2i-2k-1) \Delta \state \nonumber
\end{align}
%
is the second-order accurate upwinding solution for the polynomial interpolation. This is implemented as \href{https://github.com/robotsorcerer/levelsetpy/tree/cupy/levelsetpy/spatialderivative/upwind_first_eno2.py}{\textcolor{codelblue}{\texttt{upwindFirstENO2}}} in the \href{https://github.com/robotsorcerer/levelsetpy/tree/cupy/levelsetpy/spatialderivative}{\textcolor{codelblue}{\texttt{spatialderivative}}} folder.

To obtain a third-order accurate solution, we choose $c^\star$ as follows
\begin{align}
	c^\star = \begin{cases}
		D^3_{k^\star+1/2} &\quad \text{   if   } |D^3_{k^\star+1/2} \lowervalue| \le  |D^3_{k^\star+3/2} \lowervalue| \\
		D^3_{k^\star+3/2} \lowervalue &\quad \text{   if   } |D^3_{k^\star+1/2} \lowervalue| > |D^3_{k^\star+3/2} \lowervalue|.
	\end{cases}
\end{align}

Whence, we have
\begin{subequations}
	\begin{align}
		Q_3(\state) &= c^\star(\state-\state_{k^\star})(\state-\state_{k^\star}+1)(\state-\state_{k^\star}+2) \\
		Q_3^\prime(\state_i) &= c^\star(3(i-k^\star)^2 - 6(i-k^\star)+2) (\Delta \state)^2
	\end{align} 
\end{subequations}
for the third-order accurate correction to the approximated upwinding scheme \eqref{eq:upwinding_approx}. This is implemented as a routine in \href{https://github.com/robotsorcerer/levelsetpy/tree/cupy/levelsetpy/spatialderivative/ENO3aHelper.py}{\textcolor{codelblue}{\texttt{upwindFirstENO3aHelper}}} and called as \href{https://github.com/robotsorcerer/levelsetpy/tree/cupy/levelsetpy/spatialderivative/upwind_first_eno3.py}{\textcolor{codelblue}{\texttt{upwindFirstENO3}}} in the \href{https://github.com/robotsorcerer/levelsetpy/tree/cupy/levelsetpy/spatialderivative}{\textcolor{codelblue}{\texttt{spatialderivative}}} folder.

\subsection{HJ Weighted Essentially Nonoscillatory Solutions}
Here, we focus on weighted ENO (WENO) schemes with the same stencil as the third-order ENO scheme; however, its accuracy reaches up to fifth-order in the solution's smooth parts. Our results closely follow the presentation of Jiang and Peng in~\cite{WeightedENOPengJiang}. These WENO schemes approximate spatial derivatives at integer grid points as opposed to at half-integer grid values as we did in the ENO scheme of the previous section. 

The third-order accurate ENO scheme  essentially employs one of three substencils on a grid, namely $\{i-3, i-2, \cdots, i\}$, $\{i-2, i-1, \cdots, i+1 \}$, and $\{i-1, \cdots, i+3 \}$ on the stencils range $\{i-3, i-2, \cdots, i+3\}$. 
\begin{subequations}
	\begin{align}
		\lowervalue_{\state, i}^{-,0} &= 
		\frac{1}{3}D^+ \lowervalue_{i-3} - \frac{7}{6}D^+ \lowervalue_{i-2} + \frac{11}{6}D^+ \lowervalue_{i-1} \\
		\lowervalue_{\state, i}^{-,1} &= -\frac{1}{6}D^+ \lowervalue_{i-2} + \frac{5}{6}D^+ \lowervalue_{i-1} + \frac{1}{3}D^+ \lowervalue_{i} \\
		\lowervalue_{\state, i}^{-,2} &= -\frac{1}{3}D^+ \lowervalue_{i-1} + \frac{5}{6}D^+ \lowervalue_{i} - \frac{1}{6}D^+ \lowervalue_{i+1} 
	\end{align}
	\label{eq:substencils}
\end{subequations}
Suppose that the spatial derivative $\lowervalue_x$ is to be found using the left-leaning substencil: $\{i-3, i-2, \cdots, i \}$, then the third-order ENO scheme chooses one from \eqref{eq:substencils}
where $\lowervalue_{\state,i}^{-,p}$ denotes the third-order $p$'th substencil to $\lowervalue_{\state}(\state_i)$ for $p=0,1,2$. The WENO approximation to $\lowervalue_{\state}(\state_i)$ leverages a convex weighted average of the three substencils so that 
\begin{align}
	\lowervalue_{\state,i}^- = w_0 \lowervalue_{\state,i}^{-,0} + w_1 \lowervalue_{\state,i}^{-,1} + w_2 \lowervalue_{\state,i}^{-,2}.
\end{align}

In smooth regions of the phase space,  $w_0 = 0.1$, $w_1 = 0.6$, and $w_2=0.3$ yield the optimally accurate fifth order WENO approximation, we have for $\lowervalue_{\state,i}^-$
%
\begin{align}
	\dfrac{D^+}{30}\lowervalue_{i-3}   -& \dfrac{13}{60}D^+ \lowervalue_{i-2} + \dfrac{47}{60}D^+ \lowervalue_{i-1}   +\dfrac{9}{20}D^+ \lowervalue_{i} - 
	\dfrac{D^+}{20} \lowervalue_{i+1} \nonumber
\end{align} 
%
the fifth-order approximation $\lowervalue_{\state}(\state_i)$ and provides the smallest truncation error on a six-point stencil. 

To account for weights in non-smooth regions, however, the smoothness of the stencils \eqref{eq:substencils} can be estimated as recommended in~\cite[\S 3.4]{LevelSetsBook} so that if 
\begin{align}
	\alpha_1 = 0.1/(\sigma_1+\epsilon)^2, \, \alpha_2 = 0.6/(\sigma_2+\epsilon)^2,\, \alpha_3 = 0.1/(\sigma_3+\epsilon)^2
\end{align}
for
\begin{subequations}
	\begin{align}
		\sigma_1 &= \dfrac{13}{12}(D^+ \lowervalue_{i-3}-2D^+ \lowervalue_{i-2}+D^+ \lowervalue_{i-1})^2  \nonumber \\
		&\qquad +\dfrac{1}{4}(D^+ \lowervalue_{i-3}-4D^+ \lowervalue_{i-2}+3D^+ \lowervalue_{i-1})^2, \\
		\sigma_2 &= \dfrac{13}{12}(D^+ \lowervalue_{i-2}-2D^+ \lowervalue_{i-3}+D^+ \lowervalue_{i})^2  + \nonumber \\
		& \qquad \dfrac{1}{4}(D^+ \lowervalue_{i-2}-D^+ \lowervalue_{i})^2,\\
		\sigma_3 &= \dfrac{13}{12}(D^+ \lowervalue_{i-1}-2D^+ \lowervalue_{i}+D^+ \lowervalue_{i+1})^2  + \nonumber \\
		& \qquad \dfrac{1}{4}(3D^+ \lowervalue_{i-1}-4D^+ \lowervalue_{i}+ D^+\lowervalue_{i+1})^2,
	\end{align}
\end{subequations}
and
\begin{align}
	\epsilon&=10^{-6} \max\{D^+ \lowervalue_{i-3}, D^+ \lowervalue_{i-2}, D^+ \lowervalue_{i-1}, D^+ \lowervalue_{i} D^+ \lowervalue_{i+1}\} \nonumber \\
	& \qquad +10^{-99}
\end{align}
then, we may define the weights for the WENO scheme as 
\begin{align}
	w_1 = \alpha_1 / \sum_{i=1}^3 \alpha_i, \,w_2 = \alpha_2 / \sum_{i=1}^3 \alpha_i,\,w_3 = \alpha_3 / \sum_{i=1}^3 \alpha_i.
\end{align}
This approximates the optimal weights $w_0 = 0.1$, $w_1 = 0.6$ and $w_2=0.3$ for decently smooth $\sigma_k$ that can be dominated by $\epsilon$. Our implementation is the routine  \href{https://github.com/robotsorcerer/levelsetpy/tree/cupy/levelsetpy/spatialderivative/upwind_first_weno5a.py}{\textcolor{codelblue}{\texttt{upwindFirstWENO5a}}} which can be called as \href{https://github.com/robotsorcerer/levelsetpy/tree/cupy/levelsetpy/spatialderivative/upwind_first_weno5.py}{\textcolor{codelblue}{\texttt{upwindFirstWENO5}}}.

\subsection{Lax-Friedrichs Monotone Difference Schemes}
\label{ssec:lax-friedrichs}
We now describe a convergent monotone difference spatial approximation scheme for scalar conservation laws of the form 
\begin{align}
	\lowervalue_t + \sum_{i=1}^{N} &f_i(\lowervalue)_{x_i} = 0 \text{   for  } t>0, \state=(x_1, \cdots, x_N) \in \ren \nonumber \\
	\lowervalue(\state, 0) &= \lowervalue_0(\state), \text{ for } \state \in \ren
\end{align}

Suppose that $N=1$, let us define $\lambda_x = \Delta t/ \Delta \state$, $\Delta^+_x = \lowervalue_{j+1} - v_j$, and $\Delta^-_x = \lowervalue_{j} - \lowervalue_{j-1}$. Then at the $nth$ time step, the Lax-Friedrichs scheme is~\cite{CrandallLaxFriedrichs}
\begin{align}
	\lowervalue_j^{n+1} = \lowervalue_j^n - \dfrac{\lambda_x}{2} \Delta_x^0 f(\lowervalue_j^n) + \dfrac{1}{2} \Delta^+_x \Delta^-_x \lowervalue_j^n.
\end{align}

Furthermore, if we define the flux on the state space as 
\begin{align}
	g(\lowervalue_j, \lowervalue_{j-1}) = \dfrac{f(\lowervalue_j)+f(\lowervalue_{j-1})}{2} - \dfrac{1}{2} \lambda_x (\lowervalue_j - \lowervalue_{j-1}),
\end{align}
we may write $\lowervalue_j^{n+1} = \lowervalue_j^n - \lambda_x^+ (\lowervalue_j, \lowervalue_{j-1})$.
%

The Lax-Friedrichs scheme is monotone on the interval $[a,b]$ if the CFL condition $\lambda_x \max_{a \le \lowervalue \le b} |f^\prime(\lowervalue)| \le 1$
%
%
for $(a, b) >0$
and the upwind differencing scheme for a nondecreasing $f$ is $\lowervalue_j^{n+1} 
= \lowervalue_j^n - \lambda_x \Delta^+_x f(\lowervalue_{j-1}^n)$.
%
%
For a non-increasing $f$, we have $	\lowervalue_j^{n+1} = \lowervalue_j^n - \lambda_x \Delta_x^+ f(\lowervalue_{j}^n)$.
%
%
Our Lax-Friedrichs  implementation is implemented in the  \href{https://github.com/robotsorcerer/levelsetpy/blob/cupy/levelsetpy/explicitintegration/term/term_lax_friedrich.py}{\textcolor{codelblue}{\texttt{explicitintegration/term}}} folder. 
\section{Temporal Discretization: Method of Lines}
\label{sec:temporal}
Here, we describe further improvements on the numerical derivatives of HJ equations by further improving the fifth order accurate HJ WENO schemes presented in section \ref{sec:upwinding}. We adopt the \textit{method of lines} (MOL) used in converting the time-dependent PDEs to ODEs. 
%
%
Our presentation follows the total variation diminishing (TVD) Runge Kutta (RK) schemes with Courant-Friedrichs-Lewy (CFL) conditioning imposed for stability as presented in~\cite{ShuOsherEfficientENO} and implemented in \texttt{MATLAB}\textregistered\ in~\cite{MitchellLSToolbox}. 

\subsection{Higher-Order TVD-RK Time Discretizations}
To adopt the method of lines, the $N$-dimensional levelset representation of $\lowervalue$ is first rolled into a $1$-D vector and an adaptive integration step size, $\Delta t$, is chosen to guarantee stability following the recommendation in \cite{ShuOsherEfficientENOII}. The forward Euler algorithm thus becomes
\begin{align}
	\lowervalue(\state, t + \Delta t) = \lowervalue(\state, t) + \Delta t \Upsilon(\state, \lowervalue(\state, t))
\end{align}
where $\Upsilon$ is now the function to be integrated.

A standard MOL can then be applied for the integration similar to ODEs (we have followed Mitchell's \cite{MitchellLSToolbox} code layout to provide consistency for MATLAB users).  
We implement TVD-RK MOL schemes up to third-order accurate forward Euler integration schemes and the calling signature is as described in \autoref{lst:mol}.

\begin{minipage}{0.95\columnwidth} 
	\centering
	\begin{lstlisting}[caption={CFL-constrained method of lines routines.},label={lst:mol},style=pythonStyle, language=python]
		|\Hilight{pythoncolor\coloropacity}|  odeCFLx(schemeFunc, tspan, y0, options, schemeData)
	\end{lstlisting}
\end{minipage}
where $x$ could be one of $1, 2$, or $3$ to indicate one of  first-order, second-order, or third-order accurate TVD-RK scheme. The routine \texttt{schemeFunc} is typically one of the Lax-Friedrichs approximation routines (implemented as \href{https://github.com/robotsorcerer/levelsetpy/blob/cupy/levelsetpy/explicitintegration/term/term_lax_friedrich.py}{\textcolor{codelblue}{\textcolor{codelblue}{\texttt{termLaxFriedrichs}}}}) in the folder \href{https://github.com/robotsorcerer/levelsetpy/tree/cupy/levelsetpy/explicitintegration/term}{\textcolor{codelblue}{\texttt{explicitintegration/term}}}. It approximates the HJ equation based on dissipation functions (shortly introduced). 

The first-order accurate TVD (it is total variation bounded [TVB] actually) together with the spatial discretization used for the PDE is equivalent to the forward Euler method. We implement this as \href{https://github.com/robotsorcerer/levelsetpy/blob/cupy/levelsetpy/explicitintegration/integration/ode_cfl_1.py}{\textcolor{codelblue}{\texttt{odeCFL1}}}.

The second-order accurate TVD-RK scheme follows the RK scheme by evolving the Euler step to $t^n + \Delta t$,
\begin{align}
	\dfrac{\lowervalue^{n+1} - \lowervalue^n}{\Delta t} + F^n \cdot \nabla \lowervalue^n = 0.
\end{align}
A following Euler step to $t^n + 2 \Delta t$ follows such that 
\begin{align}
	\dfrac{\lowervalue^{n+2} - \lowervalue^{n+1}}{\Delta t} + F^{n+1} \cdot \nabla \lowervalue^{n+1} = 0
\end{align}
before a convex combination of the initial value function and the result of the preceding Euler steps is taken in the following averaging step, $\lowervalue^{n+1} = \dfrac{1}{2}\{\lowervalue^n + \lowervalue^{n+2}\}$.
%
%
The equation in the foregoing produces the second-order accurate TVD approximation to $\lowervalue$ at $t^n + \Delta t$, implemented as \href{https://github.com/robotsorcerer/levelsetpy/blob/cupy/levelsetpy/explicitintegration/integration/ode_cfl_2.py}{\textcolor{codelblue}{\texttt{odeCFL2}}}.

With the third-order accurate TVD-RK scheme, the first two advancements in forward Euler schemes are computed but with a different averaging scheme, $\lowervalue^{n+1/2} = \dfrac{1}{4}\{3\lowervalue^n + \lowervalue^{n+2}\}$
%
which averages the previous two solutions at $t^n + \dfrac{1}{2} \Delta t$. The third Euler advancement step to $t^n + \dfrac{3}{2} \Delta t$ is 
\begin{align}
	\dfrac{\lowervalue^{n+\frac{3}{2}} - \lowervalue^{n+\frac{1}{2}}}{\Delta t} + F^{n+\frac{1}{2}} \cdot \nabla \lowervalue^{n+\frac{1}{2}} = 0
\end{align}
together with the averaging scheme, $\lowervalue^{n+1} = \dfrac{1}{3}\{\lowervalue^n + 2 \lowervalue^{n+\frac{3}{2}}\}$
%
%
to produce a third-order accurate approximation to $\lowervalue$ at time $t^n + \Delta t$, implemented as \href{https://github.com/robotsorcerer/levelsetpy/blob/cupy/levelsetpy/explicitintegration/integration/ode_cfl_3.py}{\textcolor{codelblue}{\texttt{odeCFL3}}}.
\section{Numerical Validation}
\label{sec:results}
In this section, we provide a representative problem  and amend it to an HJ PDE form that can be resolved with our toolbox. 
%
We consider a \textit{collection/family of differential games}, $\Upsilon = \{\Gamma_1, \cdots, \Gamma_g\}$, where each game may be characterized as a pursuit-evasion \textit{game}, $\Gamma$. Each player in a game shall constitute either a pursuer ($\pursuer$) or an evader ($\evader$) and such a game terminates when \textit{capture} occurs. 

\subsection{The Rockets Launch Problem}
\label{subsec:rockets_launch}
We consider the rocket launch problem of Dreyfus~\cite{Dreyfus1966} and amend it to a differential game between two identical rockets, $\pursuer$ and $\evader$, on an $(x,z)$ cross-section of a Cartesian plane. We want to compute  the backward reachable tube (BRT)~\cite{Mitchell2005} of the \textit{approximate} terminal surface's boundary for a predefined target set over a time horizon (\ie the target tube).  
The BRT entails the state-space regions for which min-max operations over either \textit{strategy} of $\pursuer$ or $\evader$ is below zero, and where the \textit{variational HJI PDE} is exactly zero.  

%
%
\begin{figure}[t!]
	\centering
	\includegraphics[width=.65\columnwidth]{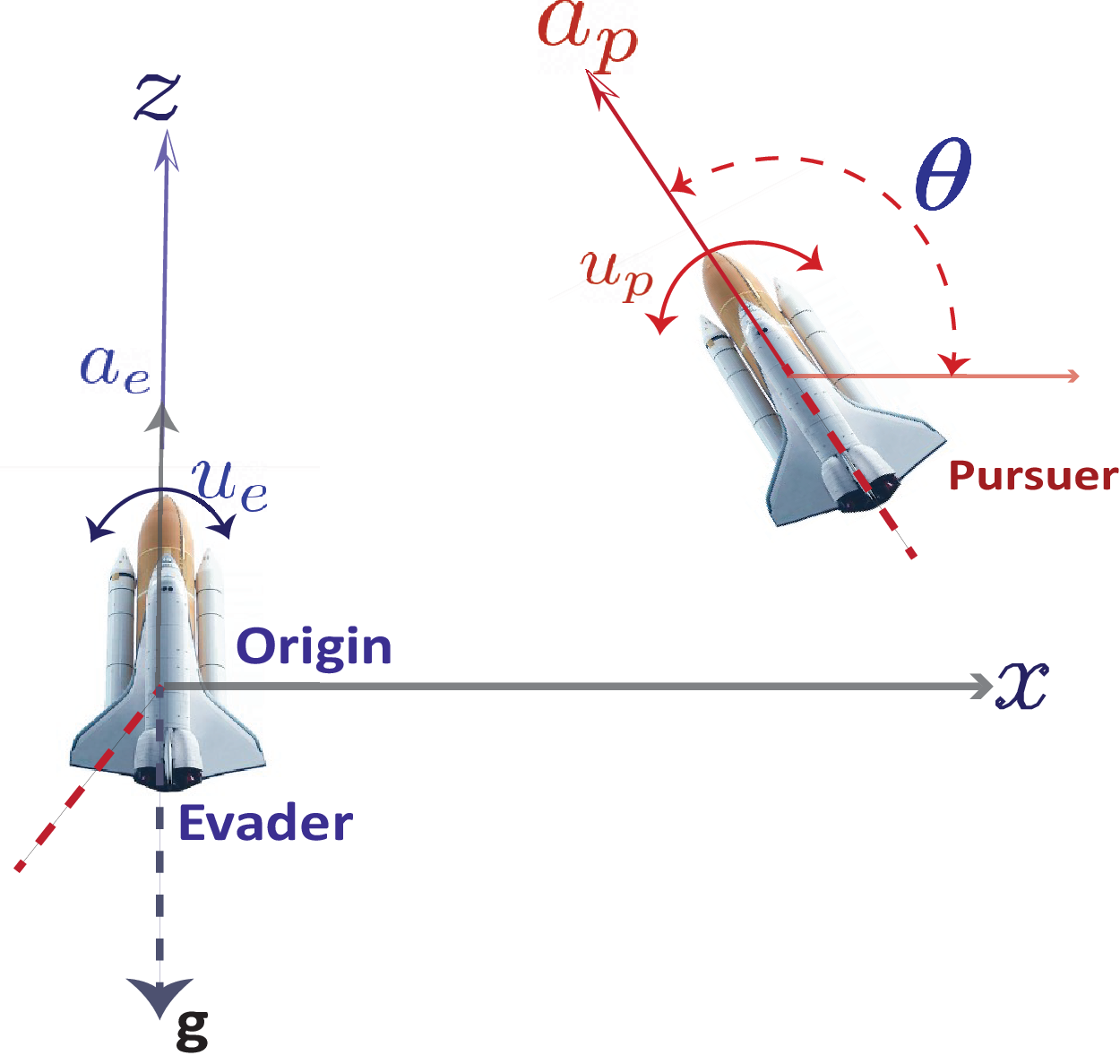}
	\caption{Motion of two rockets on a Cartesian $\bm{xz}$-plane with a thrust inclination in relative coordinates given by $\theta:=u_p- u_e$.}
	\label{fig:rocket_relative}
\end{figure}

For a two-player differential game, let $\pursuer$ and $\evader$ share identical dynamics in a general sense so that we can freely choose the coordinates of $\pursuer$; however, $\evader$'s origin is a distance $\payoff$ away from $(x,z)$ at plane's origin (see Fig. \ref{fig:rocket_relative}) so that the $\pursuer\evader$ vector's inclination measured counterclockwise from the $\state$ axis is $\theta$. 


Let the states of $\pursuer$ and $\evader$  be denoted by $(\state_p, \state_e)$. Furthermore, let the $\pursuer$ and $\evader$ rockets be driven by their thrusts, denoted by $(u_p, u_e)$ respectively  (see Figure \ref{fig:rocket_relative}). Fix the rockets' range so that what is left of the motion of either $\pursuer$ or $\evader$'s  is restricted to orientation on the $(x,{z})$ plane as illustrated in \autoref{fig:rocket_relative}. It follows that the relevant \textit{kinematic equations} (KE) (derived off Dreyfus'~\cite{Dreyfus1966} single rocket dynamics) is
\begin{subequations}
	\begin{align} 
		\dot{x}_{2e} &= x_{4e};   &\dot{x}_{2p} &= x_{4p},  \\ 
		\dot{x}_{4e} &= a \sin u_e - g; &\dot{x}_{4p} &= a \sin u_p - g
	\end{align}
	\label{eq:dreyfus_mitter_relevant_eq}
\end{subequations}
\noindent where $a$ and $g$ are respectively the acceleration and gravitational accelerations (in feet per square second) \footnote{We set $a=1 ft/sec^2$ and $g=32 ft/sec^2$ in our simulation.}.  


As long as $\evader$  remains within this target region or \textit{backward reachable tube} (or BRT), $\pursuer$ cannot cause damage or exercise an action of deleterious consequence on, say, the territory being guarded by $\evader$. Setting up $\evader$ to maximize a payoff quantity  with the largest possible margin or at least frustrate the efforts of $\pursuer$ with minimal collateral damage while the pursuer minimizes this quantity constitutes a terminal value \textit{optimal} differential game: there is no optimal pursuit without an optimal evasion. 

$\pursuer$'s motion relative to $\evader$'s  along  the $(\state,\bm{z})$ plane includes the relative orientation, the control input, shown in \autoref{fig:rocket_relative} as $\theta=u_p- u_e$. Following the conventions in \autoref{fig:rocket_relative}, the game's relative equations of motion in \textit{reduced space}~\cite[\S 2.2]{Isaacs1965} 
is $\bm{x} = (x, {z}, {\theta})$ where ${\theta }\in \left[-\frac{\pi}{2}, \frac{\pi}{2}\right)$ and $(x,z) \in \bb{R}^2$ are 
%
\begin{align}
	\dot{\state} = 
	\begin{cases}
		\dot{x} &= a_p \cos \theta + u_e x, \\
		\dot{z} &=a_p \sin \theta + a_e + u_e x - g, \\
		\dot{\theta} &= u_p -u_e.
	\end{cases}
	\label{eq:rocket_me}
\end{align}

The capture radius of the origin-centered circle $\payoff$ (we set $\payoff=1.5$ ft) is $\|\pursuer \evader\|_2 $ so that $\payoff^2 =  x^2 + z^2.$
%
All capture points are specified by  the variational HJ PDE~\cite{Mitchell2005}: 
\begin{align}
	\dfrac{\partial \payoff}{\partial t} (\bm{x},t) + \min \left[0, \hamfunc\bigg(\state, \frac{\partial \payoff(\state, t)}{\partial \state}\bigg)\right] \le 0,
\end{align}
with Hamiltonian given by
\begin{align}
	\hamfunc(\state, p) &= -\max_{u_e \in [\underline{u}_e, \bar{u}_e]} \min_{u_p \in [\underline{u}_p, \bar{u}_p].
	} \begin{bmatrix}
		p_1 & p_2 & p_3
	\end{bmatrix} 
	\nonumber \\
	& \qquad 
	\begin{bmatrix}
		a_p \cos \theta + u_e x \\
		a_p \sin \theta + a_e + u_p x - g \\
		u_p -u_e
	\end{bmatrix}.
	\label{eq:ham_def}
\end{align}
Here,  $p$ are the co-states, and $[\underline{u}_e, \bar{u}_e]$ denotes extremals that the evader must choose as input in response to the extremal controls that the pursuer plays \ie $[\underline{u}_p, \bar{u}_p]$. 
%
Rather than resort to analytical \textit{geometric reasoning}, we may analyze possibilities of behavior by either agent via a principled numerical simulation. This is the essence of this work. 
From \eqref{eq:ham_def}, set $\underline{u}_e = \underline{u}_p = \underline{u} \triangleq -1$ and $\bar{u}_p = \bar{u}_e = \bar{u} \triangleq +1$ so that $\hamfunc(\state, p)$ is
\begin{align}
	& -\max_{u_e \in [\underline{u}_e, \bar{u}_e]} \min_{u_p \in [\underline{u}_p, \bar{u}_p]
	} 
	\begin{bmatrix}
		p_1(a_p \cos \theta + u_e x) + \nonumber \\ p_2 (a_p \sin \theta + a_e + \nonumber \\ u_p x - g) + p_3 (u_p -u_e)
	\end{bmatrix}, \nonumber \\
	%
	%
	&\triangleq -a p_1 \cos \theta - p_2 (g - a - a\sin \theta) -\bar{u} |p_1 x +  p_3 | \nonumber \\ 
	& \qquad + \underline{u} |p_2 x + p_3 |,
	\label{eq:rocket_hamfunc}
\end{align}
where the last line follows from setting $a_e = a_p \triangleq a$.

For the target set guarded by $\evader$, we choose an implicitly constructed cylindrical mesh on a three-dimensional grid. The grid's nodes are uniformly spaced apart at a resolution of $100$ points per dimension over the interval $[-64, 64]$. In numerically solving for the Hamiltonian  \eqref{eq:rocket_hamfunc}, a TVD-RK discretization scheme~\cite{OsherShuENO} based on fluxes is used in choosing smooth nonoscillatory results as described in \S \ref{sec:temporal}. Denote by $(x, y, z)$ a generic point in $\bb{R}^3$ so that given mesh sizes $\Delta x, \, \Delta y, \, \Delta z, \, \Delta t \, > 0$, letters $u,v,w$ represent functions on the $x,y,z$ lattice: $\Delta=\{(x_i,y_j, z_k): i, j, k \in \bb{Z}\}$. 
%
%
%

\begin{figure*}[tb!]
	\centering
	\begin{minipage}[tb]{.31\textwidth}
		\includegraphics[width=\textwidth]{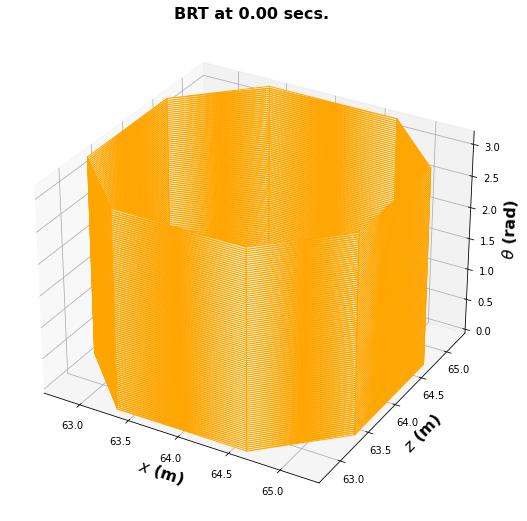}
	\end{minipage}
	\hfill
	%
	\begin{minipage}[tb]{.31\textwidth}
		\includegraphics[width=\textwidth]{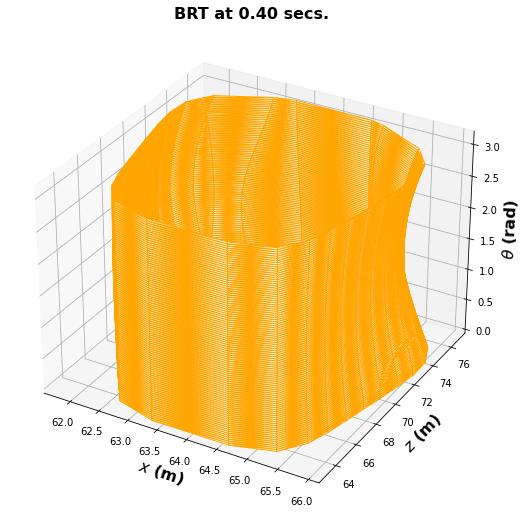}
	\end{minipage}
	\hfill
	\begin{minipage}[tb]{.31\textwidth}
		\includegraphics[width=\textwidth]{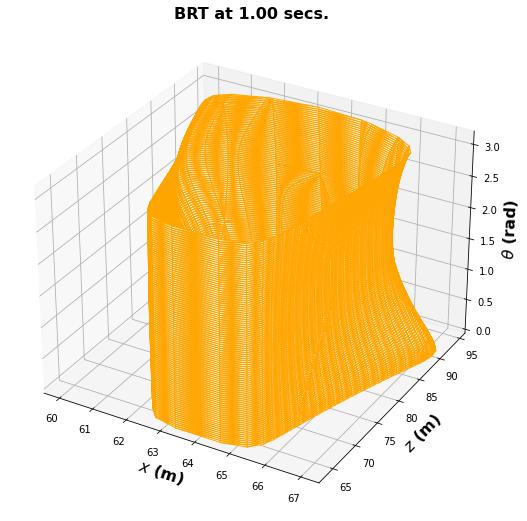}
	\end{minipage}
	\caption{\footnotesize{\textit{(Left to Right)}: Backward reachable tubes  (capture surfaces) for the rocket system (\cf \autoref{fig:rocket_relative}) optimized for the paths of slowest-quickest descent in equation \eqref{eq:ham_def} at various time steps during the differential game. 
			In all, the BRTs were computed using the method outlined in \cite{Crandall1984, LevelSetsBook, Mitchell2020}. We set $a_e = a_p = 1ft/sec^2$ and $g=32 ft/sec^2$ as in Dreyfus' original example. }}
	\label{fig:rockets_results}
\end{figure*}

\begin{minipage}{0.95\columnwidth} 
	\centering
	\begin{lstlisting}[caption={HJ ENO2 computational scheme for the rockets.}, label={lst:rockets},style=pythonStyle, language=python]
		|\Hilight{pythoncolor\coloropacity}|finite_diff_data = {"innerFunc": termLaxFriedrichs,
		|\Hilight{pythoncolor\coloropacity}|"innerData": {"grid": g, "hamFunc": rocket_rel.ham,
		|\Hilight{pythoncolor\coloropacity}|"partialFunc": rocket_rel.dissipation,
		|\Hilight{pythoncolor\coloropacity}|"dissFunc": artificialDissipationGLF,
		|\Hilight{pythoncolor\coloropacity}|"CoStateCalc": upwindFirstENO2},
		|\Hilight{pythoncolor\coloropacity}|"positive": True}  // direction of approx. growth 
	\end{lstlisting}
\end{minipage}

The Hamiltonian, upwinding scheme, flux dissipation method, and the overapproximation parameter for the ENO polynomial interpolatory data used in geometrically reasoning about the \textit{target tube} is as seen in \autoref{lst:rockets}. The data structure \texttt{finite\_diff\_data} contains all the routines needed for adding dynamics to the original implicit surface representation of $\lowervalue(\state, t)$. The monotone spatial upwinding scheme used (here \texttt{termLaxFriedrichs} described in \S \ref{ssec:lax-friedrichs}) is passed into the \texttt{innerFunc} query field. The explicit form of the Hamiltonian (see \eqref{eq:rocket_hamfunc}) is passed to the \texttt{hamFunc} query field, and the grid is passed to the \texttt{grid} field. We adopt a second-order accurate upwinding scheme together with the a \texttt{Lax-Friedrichs} conditioner for numerical stability. To indicate that we intend to overapproximate the value function, we specify a \texttt{True} parameter for the \texttt{positive} query field.

Using our GPU-accelerated toolbox, we compute the \textit{overapproximated} BRT of the game over a time span of $[-2.5, 0]$ seconds during 11 global optimization time steps (the global steps constitute the time-horizon over which the BRT is computed).  
The initial value function (leftmost inset of \autoref{fig:rockets_results}) is represented as a (closed) dynamic implicit surface over all point sets in the state space (using a signed distance function) for a coordinate-aligned cylinder whose vertical axes runs parallel to the orientation of the rockets depicted in \autoref{fig:rocket_relative}. 
The two middle capture surfaces indicate the evolution of the capture surface (here the zero levelset) of the target set upon the optimal response of the evader to the pursuer. We reach convergence at the eleventh global optimization timestep (rightmost inset of \autoref{fig:rockets_results}). The BRTs at representative time steps in the optimization procedure is depicted in \autoref{fig:rockets_results}. 

Reachability~\cite{Mitchell2001, LygerosReachability} thus affords us an ability to numerically reason about the behavior of these two rockets aforetime without closed-form geometrical analysis. To do this, we have passed relevant parameters to the package as shown in \autoref{lst:rockets} and run a CFL constrained optimization scheme (as in \autoref{lst:mol}) for a finite number of global optimization timesteps. 

\section{Comparisons and Conclusion}
\label{sec:conclude}
We compare evaluation times among our GPU-implementation, Mitchell's~\cite{MitchellLSToolbox}, and our \texttt{Numpy} CPU implementation for various problems. We refer readers to detailed problem description in this article's \href{https://scriptedonachip.com/downloads/Papers/LevPy.pdf}{journal submission version}.  Table \ref{tab:time_compute} depicts the time it takes to run the TVD-RK scheme for other reachable  problems solved with our library in comparison to Mitchell's toolbox~\cite{MitchellLSToolbox} and other implementations. The column \texttt{Avg. local} is the average time of running one single step of the TVD-RK scheme (cref{sec:temporal}) while the  \texttt{ Global } column denotes the average time to compute the full TVD-RK solution to the HJ PDE. Each time query field represents an average over 20 experiments.
\begin{table*}[tb!]
	\caption{Time to Resolve HJ PDEs. }
	\label{tab:time_compute}
	\begin{tabular}{|p{2.2cm}|c|p{1.2cm}|c|p{1.2cm}|c|p{1.2cm}|r|}
		\hline
		\multirow{2}{*}{\backslashbox{Expt}{Lib}}  & \multicolumn{2}{|c|}{\footnotesize{levelsetpy GPU Time (secs)}} & \multicolumn{2}{|c|}{\footnotesize{levelsetpy CPU Time (secs)}} & \multicolumn{2}{|c|}{\footnotesize{MATLAB CPU (secs)}} \\ 
		~  & \footnotesize{Global} & \footnotesize{Avg. local} & \footnotesize{Global} & \footnotesize{Avg. local} & \footnotesize{Global} & \footnotesize{Avg. local} \\ \hline 
		Rockets & $11.5153\pm 0.038$  & $1.1833$ & $107.84 \pm 0.42$ & $10.4023$  &  $138.50$ & $13.850$   \\ \hline
		Doub. Integ.   &  $14.7657 \pm 0.2643$  & $1.5441$ & $3.4535 \pm 0.34$ & $0.4317$ &  $5.23$ &$0.65375$ \\ \hline 
		Air 3D & $30.8654 \pm 0.1351$ & $3.0881$ & $129.1165\pm 0.13$ & $12.6373$  & $134.77$ & $16.8462$ \\ \hline
		Starlings & $8.6889 \pm 0.8323$  & $0.42853$ & $15.2693 \pm 0.167$ & $7.4387$  & N/A & N/A \\ \hline
	\end{tabular}
\end{table*}
Computation is significantly faster with our GPU implementation in all categories. In Air3D and the rockets launch problems, the average local time for computing the solutions to the stagewise HJ PDEs is an improvement of $\sim76\%$; the global time is a gain of $76.09\%$ over Mitchell's~\cite{MitchellLSToolbox}'s MATLAB scheme. Similarly, substantial computational gains are achieved for the two rockets differential game problem: $89\%$ faster global optimization time and $88.62\%$ average local computational time compared to our \text{CPU} implementations in \texttt{Numpy}. For the rockets game, we notice a speedup of almost $92\%$ in global optimization with the \texttt{GPU} library versus an $89.32\%$ gain using our \texttt{CPU}-\texttt{NUMPY} library. Notice the exception in the \texttt{Double Integrator} experiment, however: local and global computations take a little longer compared to deployments on the~\texttt{Numpy} CPU implementation and \cite{MitchellLSToolbox}'s native ~\texttt{MATLAB}\textregistered toolbox. We attribute this to the little arrays' sizes. 
Nevertheless, we still see noticeable gains in using our CPU implementation as opposed to~\cite{MitchellLSToolbox}'s~\texttt{MATLAB}\textregistered implementation. 

On a CPU, owing to efficient arrays arithmetic native to ~\cite{Numpy}'s Numpy library, the average time to compute the zero levelsets per optimization step for the \texttt{odeCFLx} functions is faster with our Numpy implementation compared against~\cite{MitchellLSToolbox} LevelSets \texttt{MATLAB\textregistered} Toolbox library across all experiments. The inefficiencies of \texttt{MATLAB\textregistered}'s array processing routine becomes pronounced in the time to finish the overall HJ PDE resolution per experiment. For \texttt{CPU} processing of HJ PDEs, it is reasonable, based on these presented data to expect that users would find our library far more useful for everyday computations in matters relating to the numerical resolution of HJ PDEs. 

HJ PDE's are increasingly becoming a useful tool in control and learning applications.  We have presented all the essential components of the python version of the \texttt{LevelSet} toolbox for numerically resolving HJ PDEs and for advancing co-dimension one interfaces on Cartesian grids. We have motivated the work presented with a numerical example to demonstrate the efficacy of our numerical implementation and we have provided a compendium of comparisons with other implementations that are available. It stands to reason that our implementation is the fastest available to the best of our knowledge. 


\addtolength{\textheight}{-3cm}   

\bibliographystyle{IEEEtran}
\bibliography{cdc24}
\end{document}